\documentclass[preprint,aps,showpacs,amsmath,amssymb]{revtex4} %

\usepackage{graphicx}
\usepackage{dcolumn}
\usepackage{bm}

\begin{document}
\draft
\title{Stopping and Time Reversal of Light in Dynamic Photonic Structures via Bloch Oscillations}
\normalsize
\author{Stefano Longhi}
\address{Dipartimento di Fisica and Istituto di Fotonica e Nanotecnologie del CNR,
Politecnico di Milano, Piazza L. da Vinci 32,  I-20133 Milan,
Italy}


%
\bigskip
\begin{abstract}
\noindent It is theoretically shown that storage and time-reversal
of light pulses can be achieved in a coupled-resonator optical
waveguide by dynamic tuning of the cavity resonances without
maintaining the translational invariance of the system. The
control exploits the Bloch oscillation motion of a light pulse in
presence of a refractive index ramp, and it is therefore rather
different from the mechanism of adiabatic band compression and
reversal proposed by Yanik and Fan in recent works [M.F. Yanik and
S. Fan, Phys. Rev. Lett. {\bf 92}, 083901 (2004); Phys. Rev. Lett.
{\bf 93}, 173903 (2004)].
\end{abstract}



\maketitle

\newpage

The possibility of dynamically control the resonant properties of
microresonator systems via small refractive index modulation
represents a promising and powerful approach for an all-optical
coherent control of light in nanophotonic structures
\cite{Yanik04a,Yanik04b,Yanik04c,Yanik05,Notomi06}. Recently,
several theoretical papers have shown that a temporal modulation
of the refractive index in photonic crystals (PCs) and
coupled-resonator optical waveguides (CROWs) can be exploited to
coherently and reversibly control the spectrum of light, with
important applications such as all-optical storage of light pulses
\cite{Yanik04a,Yanik04c}, time reversal \cite{Yanik04b} and
wavelength conversion \cite{Notomi06,Gaburro06}. The existence of
a frequency shift on the spectrum of a light pulse reflected by a
shock-wave front traveling in a PC was first pointed out by Reed
{\it et al.} \cite{Reed03a,Reed03b}, and adiabatic wavelength
conversion by simple dynamic refractive index tuning of a high-$Q$
microcavity in a PC has been numerically demonstrated in
Ref.\cite{Notomi06}. In a series of recent papers, Yanik and Fan
showed that an adiabatic and translationally-invariant tuning of
the refractive index in a waveguide-resonator system can be
exploited to stop, store and time-reverse light pulses
\cite{Yanik04a,Yanik04b,Yanik04c,Yanik05b}. The general conditions
requested to coherently stop or reverse light pulses have been
stated in Refs. \cite{Yanik05,Yanik05b}, and the possibility of
overcoming the fundamental bandwidth-delay constraint of static
resonator structures has been pointed out. The basic idea of these
previous papers is that the band structure of a
translational-invariant waveguide-resonator system can be
dynamically modified by a proper tuning the refractive index {\it
without breaking the translational invariance of the system}. For
instance, stopping a light pulse corresponds to an adiabatic band
compression process: an initial state of the system, having a
relatively wide band to accommodate the incoming pulse,
adiabatically evolves toward a final state in which the bandwidth
shrinks to zero \cite{Yanik05}. In practice, the adiabatic
evolution is attained by a slow change of the refractive index of
certain cavities forming the photonic structure
\cite{Yanik04a,Yanik04b,Yanik04c}. The condition that the dynamic
refractive index change does not break the translational
invariance of the system is important because it ensures that: (i)
the system can be described in terms of a band diagram with a
dispersion relation $\omega=\omega(k)$ relating the frequency
$\omega$ and the wave vector $k$ of its eigenmodes; (ii) different
wave vector components of the pulse are not mixed, so that all the
coherent information encoded in the original pulse
are maintained while its spectrum is adiabatically changed \cite{Yanik05,Yanik05b}.\\
In this work it is shown that a coherent and reversible control of
light in a photonic structure by dynamic refractive index change
does not necessarily require to maintain the translational
invariance of the system. We illustrate this by demonstrating the
possibility of stopping and time-reversing light pulses in a CROW
\cite{Stefanou98,Yariv99,Bayindir00} with a dynamic refractive
index gradient. In this system, light stopping and reversal is not
due to adiabatic shrinking and reversal of the waveguide band
structure, as in Refs.\cite{Yanik04a,Yanik04b}, but it is a
consequence of the coherent Bloch oscillation (BO) motion of the
light pulse induced by the index gradient. It is remarkable that,
thought temporal \cite{Sterke98,Longhi01,Sapienza03} and spatial
\cite{Peschel98,Pertsch99} BOs and related phenomena have been
studied to a great extent in several linear optical systems, they
have been not yet proposed as
an all-optical means to stop or time-reverse light pulses.\\
We consider a CROW made of a periodic array of identical coupled
optical cavities, and indicate by $\omega_n=\omega_0+\delta
\omega_n(t)$ the resonance frequency of the $n$-th cavity in the
array, where $\delta \omega_n(t)$ is a small frequency shift  from
the common frequency $\omega_0$ which can be dynamically and
externally changed by e.g. local refractive index control, as
discussed in previous works \cite{Yanik04a,Notomi06}. Practical
implementations of CROW structures have been demonstrated in
photonic crystals with coupled defect cavities
\cite{Bayindir00,Olivier01} or in a chain of coupled microrings
\cite{Poon06}. In most cases, coupled mode theory
\cite{Yariv99,Yanik04a,Christodoulides02} can be used to describe
the evolution of the field amplitudes $a_n$ in the cavities and
therefore the process of coherent light control; the results
obtained from coupled-mode theory have been shown in fact to be in
excellent agreement with full numerical simulations using
finite-difference time-domain methods (see, for instance,
\cite{Yanik04a,Yanik04b}). For our system, coupled-mode equations
read
\begin{equation}
i \frac{d a_n}{dt}=-\kappa (a_{n-1}+a_{n+1})- \delta \omega_n(t)
a_n
\end{equation}
where $\kappa$ is the hopping amplitude between two adjacent
cavities, which defines the bandwidth ($4 \kappa$) of the CROW.
Note that cavity losses are not included in Eqs.(1), however a
non-vanishing loss rate would just introduce a uniform exponential
decay in time of $a_n$ which would set a maximum limit to the
achievable delay time, as discussed in Ref.\cite{Yanik04a}. As in
Refs.\cite{Yanik04a,Yanik04b,Yanik04c,Yanik05}, field propagation
is considered at a classical level; a full quantum treatment,
which would require the introduction of noise sources in Eqs.(1)
to account for quantum noise, is not necessary for the present
analysis which deals with passive CROW structures. Contrary to
Refs. \cite{Yanik04a,Yanik04b}, we assume that the
 modulation of cavity resonances used for coherent light control is not
translational invariant, i.e. $\delta \omega_n$ depends on $n$.
Precisely, we assume that a ramp with a time-varying slope
$\alpha(t)$ is imposed to the resonances of $N$
 adjacent cavities in the CROW, leading to a site-dependent
frequency shift $\delta \omega_n(t)=n \alpha(t)$ for $1 \leq n
\leq N$ and $\delta \omega_n(t)=0$ for $n>N$ and for $n<1$. Note
that the dynamic part of the CROW which realizes stopping or time
reversal of light is confined in the region $1<n<N$, which is
indicated by a rectangular dotted box in Fig.1. The total length
of the system realizing stopping or time reversal of light is
therefore $L=Nd$, where $d$ is the distance between two adjacent
cavities. The modulation $\alpha(t)$ is assumed to vanish for
$t<t_1$ and $t>t_2$ [see Fig.1(a)], $[t_1,t_2]$ being the time
interval needed to stop or time-reverse an incoming pulse. The
switch-on time $t_1$ is chosen just after the pulse, propagating
along the CROW and coming from $n= -\infty$, is fully entered in
the dynamic part of the CROW, whereas the length $L$ is chosen
long enough to ensure that the pulse remains fully confined in the
cavities $1<n<N$ for the whole time interval $[t_1, t_2]$. Note
that, as for $t<t_1$ and $t>t_2$ the pulse propagates in the CROW
at a constant group velocity, during the time interval $[t_1,t_2]$
the pulse motion as ruled by Eqs.(1) is more involved and turns
out to be fully analogous to the motion of a Bloch particle,
within a tight-binding model, subjected to a time-dependent field
$\alpha(t)$ (see, e.g., \cite{Dunlap86,Callaway74,Hartmann04}). It
is indeed such a Bloch motion that can be properly exploited to
stop or time-reverse an incoming pulse. In fact, let us suppose
that the incoming light pulse, propagating in the forward
direction of the waveguide and coming from $n= -\infty$, has a
carrier frequency tuned at the middle of the CROW transmission
band and its spectral extension is smaller than the CROW band
width $4 \kappa$. For $t<t_1$ one can then write (see the Appendix
for technical details)
\begin{equation} a_n(t)=\int_{-\pi}^{\pi} dQ \; F(Q)
\exp(iQn+2i \kappa t \cos Q),
\end{equation}
where the spectrum $F(Q)$ is nonvanishing in a small region at
around $Q=Q_0=\pi /2$. The shape of $F(Q)$ is determined from the
excitation condition of the CROW at $n \rightarrow -\infty$ or,
equivalently, from the field distribution $a_n(t_0)$ along the
CROW at a given initial time $t=t_0 <t_1$. For instance, assuming
without loss of generality $t_0=0$, the latter condition yields
for the spectrum $F(Q)=1/(2 \pi) \sum_n a_n(0) \exp(-iQn)$ (see
the Appendix). For $t_0<t<t_1$, Eq.(2) describes a pulse which
propagates along the waveguide with a group velocity $v_g=2 d
\kappa \sin Q_0=2 d \kappa$. At time $t=t_1$, we assume that the
pulse is fully entered in the box system of Fig.1, and the
modulation of cavity resonances is then switched on. The {\it
exact} solution to Eqs.(1), which is the continuation of Eq.(2) at
times $t>t_1$, can be calculated in a closed form and reads (see
the Appendix)
\begin{equation}
 a_n(t) = \exp[i
\gamma(t) n] \int_{-\pi}^{\pi} dQ \; F(Q)  \exp[iQn+i
\theta(Q,t)],
\end{equation}
where we have set $\gamma(t)=\int_{t_1}^t dt'\alpha(t')$ and
$\theta(Q,t)=2 \kappa \int_{t_1}^{t} dt' \cos[Q+\gamma(t')]+2
\kappa t_1  \cos(Q) $. At time $t_2=t_1+\tau$, the modulation is
switched off, and for $t>t_2$ one then has
\begin{eqnarray}
a_n(t) & = & \exp(i \gamma_0 n) \int_{-\pi}^{\pi} dQ \; F(Q)
\exp[i f(Q)+i \phi(Q)] \times \nonumber
\\
 & \times &  \exp[iQn+2i \kappa t \cos(Q+\gamma_0)],
\end{eqnarray}
where we have set $\gamma_0 = \int_{t_1}^{t_2} dt \; \alpha(t)$, $
f(Q)=2 \kappa t_1 \cos Q-2 \kappa t_2 \cos (Q+\gamma_0)$, and $
\phi(Q)=\theta(Q,t_2)=2 \kappa \int_{t_1}^{t_2} dt \; \cos
[Q+\gamma(t)]$.
\begin{figure}
\includegraphics[scale=0.5]{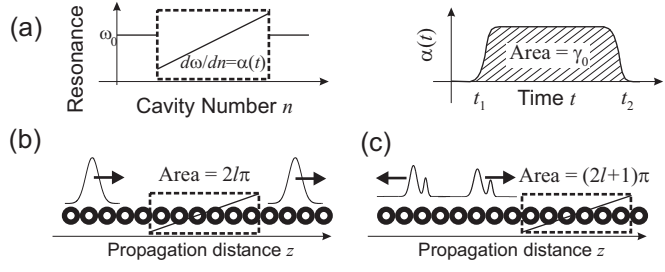} \caption{
Schematic of a dynamic CROW with a linear gradient of resonances.
(a) Distribution of cavity resonances (left) and temporal behavior
of gradient amplitude $\alpha(t)$ (right); (b) Process of pulse
storage; (c) Process of time reversal.}
\end{figure}
\noindent The modulation parameters are chosen to either store or
time reverse the incoming pulse (see Fig.1). In both cases, we
assume that  the length $L$ of the system is large enough to
entirely contain the pulse in the whole interval $[t_1,t_2]$. In
case of pulse storage, after the modulation is switched off the
pulse escapes from the system in the forward direction with the
same group velocity $v_g=2 d \kappa$ as that of the incoming
pulse, but it is delayed by a time $\sim \tau$ [see Fig.1(b)]. In
case of time reversal [Fig.1(c)], the incoming pulse is
reflected from the system, which thus acts as a phase-conjugation mirror.\\
\\
Consider first the process of {\it pulse storage}. To this aim,
let us assume that the {\it area} $\gamma_0$ be an integer
multiple of $2 \pi$. In this case, for $t>t_2$ from Eq.(4) one
obtains
\begin{equation}
a_n(t) =  \int_{-\pi}^{\pi} dQ \; F(Q) \exp [i \phi(Q)]
\exp[iQn+2i \kappa (t-\tau) \cos Q ].
\end{equation}
A comparison of Eqs.(2) and (5) clearly shows that, if $\phi=0$
the effect of the modulation is that of storing the pulse for a
time $\tau=t_2-t_1$ {\it without introducing any distortion}: in
fact, one has $a_n(t_2)=a_n(t_1)$.
\begin{figure}
\includegraphics[scale=0.4]{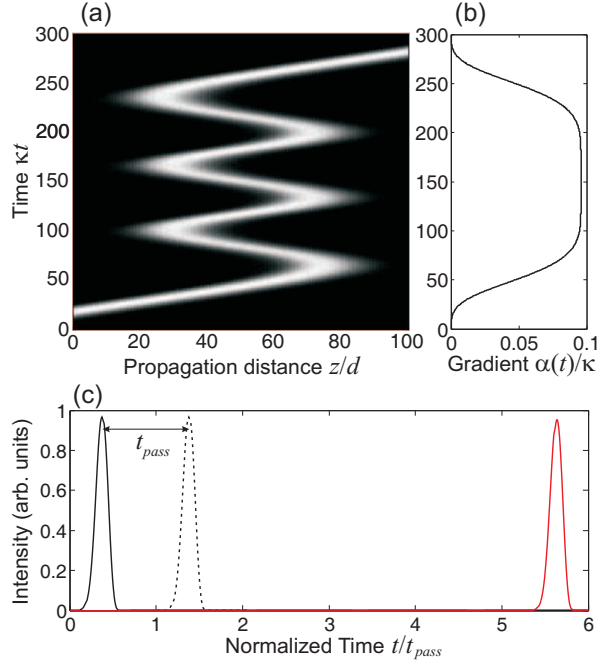} \caption{(color online) Storage of a
Gaussian pulse in a coupled resonator waveguide consisting of
$N=100$ cavities. (a) Gray-scale plot showing the space-time pulse
intensity evolution (note the characteristic BO motion). (b)
Profile of the applied modulation $\alpha(t)=\alpha_0 \exp \{
-[(\kappa t-150)/ \tau_0]^6 \}$ with $\alpha_0 / \kappa=0.0958$
and $\tau_0=106$, corresponding to an area $\gamma_0=6 \pi$. (c)
Process of pulse storage: the solid black curve is the intensity
profile of the incoming pulse as recorded in the first cavity of
the waveguide ($z=0$), whereas the solid red curve and dashed
black curve are the intensity profiles of the outcoming pulse as
recorded in the last cavity of the waveguide ($z=L=100d$) in the
presence and in the absence of the modulation, respectively. In
(c) time is normalized to the transit time $t_{pass}=N/(2
\kappa)=50/ \kappa$ of the pulse in the system.}
\end{figure}
\begin{figure}
\includegraphics[scale=0.4]{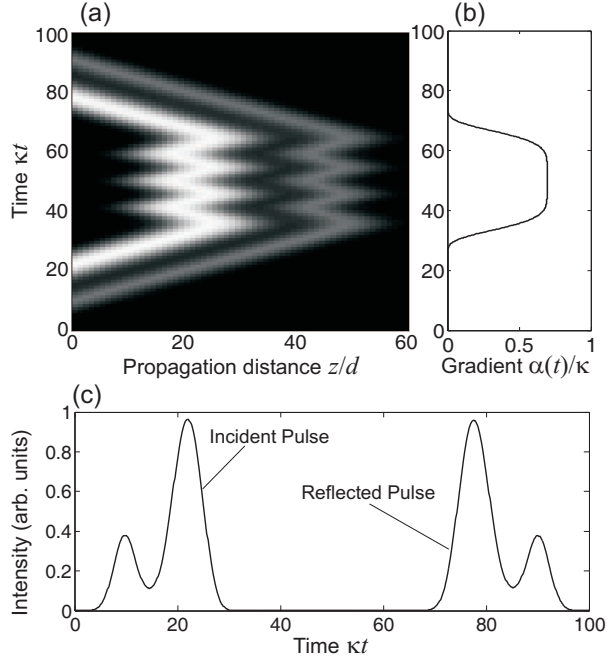} \caption{Time reversal of a
non-symmetric double-peaked optical pulse in a coupled resonator
optical waveguide consisting of $N=60$ cavities. (a) Gray-scale
plot showing the space-time pulse intensity evolution. (b) Profile
of the applied modulation $\alpha(t)=\alpha_0 \exp \{ -[(\kappa
t-50)/ \tau_0]^6 \}$ with $\alpha_0 / \kappa=0.7$ and $\tau_0=17$,
corresponding to an area $\gamma_0=7 \pi$. (c) Pulse intensity
profile versus scaled time $\kappa t$ as recorded in the first
cavity of the waveguide ($z=0$).}
\end{figure}
\noindent If the area in an integer multiple of $2 \pi$ but $\phi
\neq 0 $, the additional phase $\phi(Q)$ may introduce a
non-negligible pulse distortion. The distortionless condition
$\phi=0$ is exactly satisfied in two important cases: a step-wise
modulation $\alpha(t)=\alpha_0$ const (with $\alpha_0 \tau=2 \pi
l$, $l$ is an integer), and a sinusoidal modulation
$\alpha(t)=\alpha_0 \cos(\Omega t)$, with $\tau \Omega=2 \pi l$
and $J_0( \alpha_0/ \Omega)=0$. These two cases realize the
well-known dc or ac BO motion
\cite{Dunlap86,Callaway74,Hartmann04} of the light pulse in the
interval $[t_1,t_2]$: pulse storage is therefore due to the
periodic motion of the light pulse which returns to its initial
position after each BO (or ac field) period. It is worth
commenting more deeply the very different mechanisms underlying
light stopping in the translational-invariant system of
Ref.\cite{Yanik04a} with the one considered in the present work.
In Ref.\cite{Yanik04a}, the modulation of cavity resonances
preserves the translational symmetry and, as a consequence, cross
talk between different wave vector components of the pulse is
prevented as the waveguide band accomodating the pulse shrinks to
zero and the pulse group velocity adiabatically decreases.
Additionally, the tuning process must be slow enough to ensure
reversibility, i.e. to ensure the validity of the adiabatic
theorem. In the present work, the tuning of the cavity resonances
breaks the translational symmetry of the system and the different
wave vector components $Q$ of the pulse undergo a drift motion in
the reciprocal space according to the well-known 'acceleration
theorem' of a Bloch particle studied in solid-state physics
\cite{Callaway74} (see the Appendix for more details). The motion
of $Q$ in the reciprocal space is accompanied by a shift of the
pulse carrier frequency, which spans in a periodic fashion the
full band of the waveguide, and by a periodic motion of the pulse
in the "stopping box" of Fig.1, which is hence trapped inside it
[see Fig.2(a) to be discussed later]. In particular, for a
step-wise modulation $\alpha(t)$ which will be mainly considered
in this work \cite{note1}, the temporal periodicity of the motion
is $\tau_B=2 \pi / \alpha_0$. This is the well-known periodic
Bloch motion which is related to the existence for Eqs.(1) of a
discrete Wannier-Stark ladder spectrum instead of a continuous
band spectrum (for more details see, for instance,
\cite{Callaway74}). Therefore, as in the translational-invariant
waveguide system of Ref.\cite{Yanik04a} light stopping is achieved
by adiabatically shrinking to zero the band of the pulse, in our
system light stopping can be viewed as a trapping effect due to
the appearance of the periodic Bloch motion in the dynamic part of
the CROW structure. Note that, as opposed to the method of
Ref.\cite{Yanik04a}, in our case adiabaticity of the tuning
process is not required, however the stopping time $\tau$ is
quantized since it must be an integer multiple of the Bloch period
$\tau_B$. Nevertheless, with a suitable choice of the gradient
$\alpha_0$ (and hence of $\tau_B$), a target delay time $\tau$ can
be achieved. If $\tau_p$ is the duration of the incoming pulse to
be delayed (with $ \tau_p < 1/ \kappa$), we can estimated the
minimum length $L$ of the system as $L=L_p+L_b$, where $L_p \simeq
\tau_p v_g= 2 d \kappa \tau_p$ is the spatial extension of the
pulse in the waveguide in the absence of the modulation and $L_b
\simeq 4 \kappa d / \alpha_0$ is the amplitude of the BO motion.
Hence the minimum number of cavities of the system is given by
$N=L/d \simeq 2 \kappa (\tau_p+ 2 / \alpha_0)$. It should be noted
that in practice a sharp step-wise modulation can never be
realized, and a finite rise time during switch on and off should
be accounted for. Though $\phi(Q)$ does not exactly vanish in this
case, dispersive effects can be kept however at a small level. As
an example, Fig.2 shows the process of light storage as obtained
by a direct numerical simulation of Eqs.(1) using a super-Gaussian
profile for the gradient $\alpha(t)$. The area $\gamma_0$ is
chosen to be $6 \pi$, so that pulse trapping corresponds to three
BO periods, as clearly shown in the space-time plot of Fig.2(a).
Note that the system comprises $N=100$ cavities, and therefore the
length $L$ of the waveguide needed to perform  light storage is
$L=Nd=100d$. The intensity profile of the nearly Gaussian-shaped
incoming pulse  in the initial cavity ($z=0$) is indicated by the
black solid line in Fig.2(c). In the figure, the intensity profile
of the outcoming pulse at the last cavity of the system ($z=100
d$) is shown by the red solid line, whereas the dashed curve
indicates the intensity profile of the output pulse at the last
cavity in the absence of index gradient, i.e. when the pulse
freely propagates along the system at the group velocity $v_g=2
\kappa d$. In Figs.2(a) and 2(b), time is normalized to $1/
\kappa$, whereas in Fig.2(c) time is normalized to the transit
time $t_{pass}=L/v_g=N/(2 \kappa)$ of the pulse in the system.
Note that the maximum frequency shift of cavity resonance needed
to achieve the process of pulse storage is $\delta \omega_{max} =
\pm (N/2) \alpha_0 \sim  \pm 5 \kappa$. Assuming that a change
$\delta n$ of the refractive index $n$ produces a change $\delta
\omega \sim \omega_0 (\delta n /n)$ of the cavity resonance
$\omega_0$, the index ramp of Fig.2(b) thus corresponds to a
maximum refractive index change $\delta n / n \sim 5 \kappa /
\omega_0$. This value is comparable to the one requested for light
stopping by means of adiabatic band compression in the
translational-invariant case \cite{Yanik04a}. To get an idea of
typical values in real physical units, let us assume e.g. a
carrier angular frequency $\omega_0 \simeq 1.216 \times 10^{15}$
rad/s (corresponding to a wavelength $\lambda \simeq 1.55 \;
\mu$m) and a maximum index change $\delta n /n \sim 5 \times
10^{-4}$, which is comparable to the one used in previous studies
(see, e.g. \cite{Yanik04a,Notomi06}). The bandwidth $2 \kappa$ of
the waveguide and the transit time $t_{pass}$ in the figure are
then given by $2 \kappa  \sim 2 \times 10^{-4} \omega_0 \sim 2.43
\times 10^{11}$ rad/s (i.e. $\sim 39$ GHz) and $t_{pass}=N/(2
\kappa) \simeq 410$ ps, respectively. For such parameter values,
Fig.1 simulates the stopping of a $\sim 68$
ps-long (FWHM) Gaussian pulse with a storage time $\tau \sim 1.75$ ns.\\
\\
The process of {\it time-reversal} of a light pulse is simply
achieved when the {\it area} $\gamma_0$ is equal to $\pi$, apart
from integer multiples of $2 \pi$. In fact, in this case for
$t>t_2$ from Eq.(4) one obtains
\begin{eqnarray}
a_n(t) & = &  (-1)^n \int_{-\pi}^{\pi} dQ \; F(Q) \exp [i \phi(Q)]
\times \nonumber \\
& &  \exp[iQn-2i \kappa (t-t_1-t_2) \cos Q ],
\end{eqnarray}
A comparison of Eqs.(2) and (6) clearly shows the sign reversal of
the frequency $2 \kappa \cos Q$ for any wave number $Q$ in the
integral term, which is the signature of time reversal of the
pulse. Physically, time reversal is due to the fact that, for the
$\pi$ area, the spectrum of the wave packet in the reciprocal $Q$
space  (quasi-momentum) is shifted in the Brillouin zone from
$Q_0=\pi/2$ to $Q_0=-\pi/2$, thus producing spectral inversion. In
addition, since the group velocity is correspondingly reversed,
the pulse is reflected by the system and thus propagates backward.
As in the previous case, the process of time reversal does not
introduce pulse distortion provided that $\phi=0$. For a step-wise
modulation, contrary to the $ 2 \pi$ area case $\phi(Q)$ does not
vanish and one has $\phi(Q)=-(4 \kappa / \alpha_0) \sin Q$.
However, this additional phase term may be kept small by choosing
e.g. a sufficiently large value of $\alpha_0$, thus minimizing
pulse distortion. An example of time reversal of an asymmetric
pulse with minimal distortion, as obtained by a direct numerical
simulation of Eqs.(1) using a super-Gaussian profile for the
gradient $\alpha(t)$, is shown in Fig.3. Note that in this case
the pulse undergoes a semi-integer
number of BO periods.\\
In conclusion, it has been theoretically shown that storage and
time-reversal of light can be realized by exploiting BOs in a
dynamic coupled-resonator waveguide. The proposed scheme is rather
distinct from the adiabatic band compression technique recently
proposed in Refs. \cite{Yanik04a,Yanik04b}, and provides a
noteworthy example of coherent light control in a system with
broken translational invariance.

\appendix
\section{}
In this Appendix we provide a detailed derivation of the solution
to the coupled-mode equations (1) at times $t<t_1$, $t_1<t<t_2$
and $t>t_2$ presented in the text [Eqs.(2), (3) and (4)]. To this
aim, we follow a rather standard technique (see, for instance,
\cite{Dunlap86}) and introduce the time-varying Fourier spectrum
$G(Q,t)$ defined by the relation
\begin{equation}
G(Q,t)=\frac{1}{2 \pi} \sum_{n=-\infty}^{\infty}a_n(t) \exp(-iQn).
\end{equation}
The amplitudes $a_n(t)$ can be derived from the spectrum $G(Q,t)$
after inversion according to the relation
\begin{equation}
a_n(t)=\int_{-\pi}^{\pi} dQ \; G(Q,t) \exp(iQn).
\end{equation}
Using the coupled-mode equations (1) with $\delta
\omega_n(t)=\alpha(t) n$, the following differential equation for
the spectrum $G$ can be easily derived \cite{note2}
\begin{equation}
\frac{\partial G}{\partial t} + \alpha \frac{\partial G}{\partial
Q}= 2 i \kappa G \cos Q.
\end{equation}
For $t<t_1$, we have $\alpha(t)=0$, and therefore the solution to
Eq.(A3) is simply given by
\begin{equation}
G(Q,t)=F(Q) \exp(2 i \kappa t \cos Q) \; \; (t<t_1),
\end{equation}
where the profile $F(Q)$ is determined by the spectrum $G$ at an
initial time $t=t_0$ by means of Eq.(A1) once the field
distribution $a_n(t_0)$ is assigned. Note that substitution of
Eq.(A4) into Eq.(A2) yields Eq.(2) given in the text.\\
For $t>t_1$, $\alpha(t)$ is nonvanishing and the solution to
Eq.(A3), which is a continuation of Eq.(A4) for times $t>t_1$, can
be easily obtained after the change of variables $\eta=t$ and
$\xi=Q-\gamma(t)$, where we have set
\begin{equation}
\gamma(t) =\int_{t_1}^t dt' \alpha(t').
\end{equation}
With these new variables, Eq.(A3) is transformed into the equation
\begin{equation}
\frac{\partial G(\xi, \eta)} {\partial \eta}= 2 i \kappa G(\xi,
\eta) \cos[\xi+\gamma(\eta) ],
\end{equation}
 which can be easily integrated with the initial condition $G(\xi, \eta=t_1)=
 F(\xi) \exp(2i \kappa t_1 \cos \xi)$. Upon re-introducing the old variables $Q$ and $t$,
one then obtains
\begin{widetext}
\begin{equation}
G(Q,t)=F(Q-\gamma(t)) \exp[2 i \kappa t_1 \cos (Q-\gamma(t))] \exp
\left\{2 i \kappa \int_{t_1}^t dt' \;\cos[Q+\gamma(t')-\gamma(t)]
\right\} \; \; ( t > t_1).
\end{equation}
\end{widetext}
Substituting  Eq.(A7) into Eq.(A2), after the change of
integration variable $Q'=Q-\gamma(t)$ and taking into account the
$2 \pi$-periodicity of the spectrum $G(Q,t)$ with respect to the
variable $Q$, one then readily obtains Eq.(3) given in text. Note
that $|G(Q,t)|^2=|F(Q-\gamma(t))|^2$, i.e. the role of the
gradient $\alpha(t)$ is to induce a rigid drift of the initial
spectrum, a result which is known as "acceleration theorem"  in
the solid-state physics context \cite{Callaway74}. In particular,
for a constant gradient $\alpha(t)=\alpha_0$, the drift of the
spectrum is uniform in time. In this case, from Eq.(A5) it follows
that, after a time $\tau_B=2 \pi / \alpha_0$ from the initial time
$t=t_1$, one has $\gamma(t_1+\tau_B)=2 \pi$ and, from Eq.(A7),
$G(Q,t_1+\tau_B)=G(Q,t_1)$, i.e. the initial field distribution in
the CROW structure is retrieved: $\tau_B$ plays the role of the BO
period which is
determined by the gradient $\alpha_0$.\\
For $t>t_2$, one has $\alpha(t)=0$ and the spectrum $G(Q,t)$ is
still given by Eq.(A7), where according to Eq.(A5) one has
$\gamma(t)=\int_{t_1}^{t_2} dt' \alpha(t') \equiv \gamma_0$ for
$t>t_2$. Note that for $t>t_2$ Eq.(A7) can be cast in the
following form
\begin{equation}
G(Q,t)=G(Q,t_2) \exp[2 i \kappa (t-t_2) \cos Q] \; \; (t>t_2)
\end{equation}
Substitution of Eq.(A8) into Eq.(A2), with $G(Q,t_2)$ given by
Eq.(A7) with $t=t_2$, and after the change of variable
$Q'=Q-\gamma_0$ in the integral of Eq.(A2), one finally obtains
Eq.(4) given in the text for the solution at times $t>t_2$.


\begin{thebibliography}{99}
\bibitem{note2} In deriving Eq.(A3)  we assumed that the relation $\omega_n=n
\alpha(t)$ holds even for $n<1$ and $n>N$, though in practice the
dynamic tuning of cavity resonances is applied solely for the
cavities with index $n$ in the range $1<n<N$ (see Fig.1). However,
taking into account that $\alpha(t)$ vanishes outside the interval
$[t_1,t_2]$ and that the size $N$ of the stopping box is large
enough to fully contain the pulse during the delay interval, i.e.
$a_n(t) \simeq 0 $ for $n<1$, $n>N$ in the time interval
$t_1<t<t_2$, the previous assumption does not change the solution
of the problem.
\end{thebibliography}
\end{document}